\documentclass[prd,showpacs,amsmath,amssymb,twocolumn,preprintnumbers,nofootinbib]{revtex4}

\usepackage[utf8]{inputenc}
\usepackage{graphicx}
\usepackage{amsmath}
\usepackage{hyperref}

\normalbaselines
\newcommand{\cqg}{Classical Quantum Gravity\ }

\newcommand{\D}{\nabla}
\renewcommand{\d}{\partial}
\renewcommand{\Re}{\mathop{\rm Re}\nolimits}
\renewcommand{\Im}{\mathop{\rm Im}\nolimits}

\begin{document}

\title{Self-interaction in the Bopp-Podolsky electrodynamics: Spacetimes with angular defects}

\author{Alexei E. Zayats}
\email{Alexei.Zayats@kpfu.ru} \affiliation{Department of
General Relativity and Gravitation, Institute of Physics, Kazan
Federal University, Kremlevskaya str. 18, Kazan 420008, Russia}

\begin{abstract}
 We consider the self-interaction phenomenon in the framework of the Bopp-Podolsky electrodynamics.
 In the present paper, we obtain the self-interaction potential energy of a charge at rest for the spacetimes with topological defects of two types: for the axially symmetric spacetime of the straight cosmic string and the spherically symmetric global monopole spacetime. It is shown that the behavior of this expression depends essentially on the angular defect, in spite of the Bopp-Podolsky model parameter, which plays the role of a scale factor. In contrast with the usual Maxwell electrodynamics, the self-interaction energy for the Bopp-Podolsky electrodynamics appears to be finite everywhere and the standard renormalization procedure is not required.
 \end{abstract}

\pacs{04.20.-q, 04.40.-b, 03.50.-z}

\maketitle

\section{Introduction}

For the case of the flat spacetime, the phenomenon of self-interaction for electric charges has been
thoroughly elaborated and described in detail (see, e.g., textbooks \cite{SFbook1,SFbook2,SFbook3}).
Based on the symmetry properties of the Minkowski spacetime it is evident to see that the self-force acting on a charge $q$ both at rest and in uniform motion has to be equal to zero. A nontrivial result can appear, only when a particle moves nonuniformly. In this case, the self-force $f_{\rm sf}$ is defined by the well-known Abraham-Lorentz-Dirac formula
\begin{equation}\label{ALD}
  f^i_{\rm sf}=\frac{2q^2}{3}\ddot{u}^k\left(\delta^i_k-u^iu_k\right).
\end{equation}
In curved spacetimes and/or in spacetimes with a nontrivial topological structure the situation becomes more sophisticated and therefore more interesting, since even for a static electric charge the self-force can be nonvanishing (see, e.g., \cite{3}).

As it is well-known, the derivation of the expression for the self-force in the Maxwell electrodynamics is inseparably linked with some kind of regularization procedure for the energy of interaction between a charge and its electromagnetic field \cite{SFbook1,SFbook2,SFbook3}. However, if we will use a suitable modification of the Maxwell theory, the problem with divergences and renormalization can be eliminated. Among these generalizations one can mark out two of the most famous: the Born-Infeld model \cite{BornInfeld} and the Bopp-Podolsky model \cite{Bopp,Podolsky1}. In contrast to the Born-Infeld model (behavior of the fields near the worldline and the self-force problem were discussed in \cite{PerlickBI}), which is a nonlinear one, the second model
suggested by Fritz Bopp and Boris Podolsky inherits linearity, but contains an additional high-derivative term
$-\frac{1}{8\pi\mu^2}\,\D_iF^{im}\,\D^kF_{km}$ (a factor $1/\mu^2$ is a parameter of the model with the dimensionality of area).

In the framework of the Bopp-Podolsky (or BP for short) electrodynamics, the potential of resting point charge and its field energy appear to be {\it finite}, at least in the Minkowski spacetime \cite{Bopp,Podolsky1,Lande2}.  Moreover, as it was proved in \cite{GPT}, even if the charge moves, there is a wide class of worldlines in the flat space for which the potential is finite (the case of uniformly accelerated motion was earlier discussed in \cite{Me}).

This remarkable feature of the BP model permits us to believe that the self-interaction energy expression will be regular for the static point charge in curved spacetimes as well. In order to state this assumption, in the present paper we consider two specific examples of non-Minkowski spacetime, which admit the derivation of the energy formula by direct calculation. Both of these spacetimes possess topological defects. The first example is the static axially symmetric spacetime of a straight cosmic string \cite{string1}.  The second spacetime has the spherical symmetry and corresponds to a global gravitating monopole \cite{monopole}. For both configurations, the self-interaction energy will be found explicitly and its behavior depending on the distance and the BP model parameter $\mu$ will be studied in detail. In addition, we will demonstrate explicitly that all obtained expressions are regular and a renormalization procedure is not required.

This paper is organized as follows. In Sect.~\ref{BPformalism}, we briefly formulate a general formalism of the Bopp-Podolsky electrodynamics and derive basic formulas for the self-interaction energy in the static case. In Sect.~\ref{axial} and \ref{spherical}, we obtain expressions for the energy when a charge is at rest in the cosmic string spacetime and the global monopole spacetime, respectively. Sect.~\ref{conclusion} is for conclusion.

\section{Bopp-Podolsky electrodynamics}\label{BPformalism}

\subsection{Basic formalism}

The formalism of BP electrodynamics starts from the Lagrangian
\begin{equation}\label{action}
    {\cal L}_{BP}=\frac{1}{16\pi}F_{ik}F^{ik}-\frac{1}{8\pi\mu^2}\,\D_iF^{im}\,\D^kF_{km}+A_ij^i\,,
\end{equation}
where $A_i$ is the electromagnetic field potential, $F_{ik}=\D_iA_k-\D_kA_i$ is the Maxwell tensor, $\nabla_i$ denotes the covariant derivative, and $j^i$ is the current density. The variation procedure with respect to the potentials $A_i$ gives the electromagnetic field equation
\begin{gather}
\D_m\left(F^{mi}+\frac{1}{\mu^2}\,G^{mi}\right)=4\pi j^i\,,\nonumber\\
G_{mn}\equiv
\D_m\D^kF_{kn}-\D_n\D^kF_{km}. \label{Meq}
\end{gather}
It is easy to see the BP Lagrangian (\ref{action}) and the BP equation (\ref{Meq}) reduce to the standard Maxwellian form when $\mu\to\infty$.

Solutions of the fourth-order equations (\ref{Meq}) can be written as a difference
between $A_i^{(0)}$ and $A_i^{(\mu)}$ (see~\cite{Bopp}),
$$A_n=A_n^{(0)}-A_n^{(\mu)},$$
where the first term is a solution to the Maxwell equation, while
the second one obeys the Proca equation with the same current density:
\begin{gather}
\D^mF_{mn}^{(0)}=4\pi j_{\,n},\label{Meq2}\\
\D^mF_{mn}^{(\mu)}+\mu^2A_n^{(\mu)}=4\pi
j_{\,n}.\label{Peq}
\end{gather}
Thus, Eq.~(\ref{Meq}) splits into two
second-order equations describing, correspondingly, massless and
massive (with the mass $\mu$) vector fields. In this sense, the BP model
has a close relationship with the Pauli-Villars regularization procedure (see, e.g., \cite{Kvasnica}) and Lee-Wick
model \cite{LeeWick} (see also the recent papers \cite{Barone}).

The electromagnetic field energy-momentum tensor derived by the variation with respect to the
metric from the action with the Lagrangian (\ref{action}) of this
model takes the form
\begin{align}
T_{ik}^{BP}&=\frac{1}{4\pi}\left(\frac14 F_{mn}F^{mn}
    g_{ik}-F_{im}{F_k}^m\right)\nonumber\\
{}&+\frac{1}{4\pi\mu^2}\left(\frac{1}{2}\,g_{ik}\,\D_pF^{pm}\,\D^qF_{qm}
    -\D^mF_{mi}\D^nF_{nk}\right.\nonumber\\
    {}&+\frac{1}{2}\,G_{mn}F^{mn}g_{ik}-G_{im}{F_k}^m-G_{km}{F_i}^m\biggr).\label{TEI0}
\end{align}
When the potential $A_i$ satisfies the BP model equation (\ref{Meq}), the energy-momentum tensor is also splitting into two parts: $T_{ik}^{BP}=T_{ik}^{(0)}-T_{ik}^{(\mu)}$,
where the first one,
\begin{gather}
    T_{ik}^{(0)}=\frac{1}{4\pi}\left(\frac14 F_{mn}^{(0)}F^{mn(0)}
    g_{ik}-F_{im}^{(0)}{F_k}^{m(0)}\right),\nonumber
\end{gather}
corresponds to the massless vector field $A_i^{(0)}$, and the second one,
\begin{align}
    T_{ik}^{(\mu)}&=\frac{1}{4\pi}\left(\frac14 F_{mn}^{(\mu)}F^{mn(\mu)}
    g_{ik}-F_{im}^{(\mu)}{F_k}^{m(\mu)}\right.\nonumber \\
    {}&+\mu^2\left[A_i^{(\mu)}A_k^{(\mu)}-\frac{1}{2}g_{ik}A_m^{(\mu)}A^{m(\mu)}\right]\biggr),
\end{align}
is the energy-momentum tensor of the Proca-type massive field
$A_i^{(\mu)}$. This field has a
negative energy and therefore it can be interpreted as a {\it phantom} one.

\subsection{Static point electric charge field in the Minkowski spacetime}

When the particle is at rest, its current density is
\begin{equation}
    j^{\,k}=q\,\delta_0^k\delta({\bf r}-{\bf r}_c),
\end{equation}
where $q$ is a charge of the particle, $\delta(x)$ is the Dirac
delta function, ${\bf r}_c$ denotes a radius-vector of a charge position, and the field potential in the framework of the BP electrodynamics \cite{Bopp,Lande2,Podolsky1} takes the form
\begin{equation}\label{statpoten}
    A_0=\frac{q}{|{\bf r}-\mathbf{r}_c|}(1-{\rm e}^{-\mu |\mathbf{r}-\mathbf{r}_c|})\,.
\end{equation}
For $\mu |\mathbf{r}-\mathbf{r}_c|\gg 1$ this formula turns into the
well-known expression for the Coulomb electrostatic potential, but at the origin it has no singularity,
$\lim\limits_{\mathbf{r}\to\mathbf{r}_c}A_0=\mu q$. The energy of
the electrostatic field (\ref{statpoten}) in the BP model is also
finite and expressed by the formula
\begin{equation}
    U_{0}=\int dV\, T_{00}=\frac{1}{2}\,\mu\,q^2.\label{E0}
\end{equation}
%The obtained value can be identified as an electromagnetic
%component of the mass of the rest charged particle, or, for
%simplicity, its ``electromagnetic'' mass.

\subsection{Self-force}

Since in the BP model the
expression of the point charge field potential is free from a
singularity at the origin, in order to derive a formula for the self-force
$f^i_{\rm sf}$, we could write down directly
\begin{equation}
f^i_{\rm sf}=q F^{ik}u_k,\label{selfforcedef}
\end{equation}
where $F^{ik}$ is the Maxwell tensor of the electromagnetic field
produced by the moving particle, and $u_k$ is its
velocity vector, and both quantities have to be calculated at the present
position of the charged particle.

However, for the static field potential (\ref{statpoten}) in the Minkowski spacetime, when $\mathbf{r}\to\mathbf{r}_c$ we have
\begin{equation}
{F}_{i0}\propto \frac{x^i-x_c^i}{|\mathbf{r}-\mathbf{r}_c|}. \label{hedge}
\end{equation}
It is easy to see that the strength tensor field (\ref{hedge}) has a hedgehog-like singularity at the point of charge. Therefore, in order to calculate the force acting on the charge we should average the expression (\ref{hedge}) over a solid angle. After this procedure we come to the predictable result: for the static charge in the Minkowski spacetime $f_{\rm sf}^i=0$. So, when we use the formula (\ref{selfforcedef}), the mentioned procedure of averaging should be always implied.

On the other hand, the expression (\ref{selfforcedef}) can be rewritten in the static case ($u^i=u^0\delta^i_0$) as follows
\begin{align}
f_i^{\rm sf}&=q\,\lim\limits_{\mathbf{r}\to\mathbf{r}_c}\d_i\left(A_{0}^{(0)}-A_0^{(\mu)}\right)u^0\nonumber\\
{}&=4\pi q^2\,\lim\limits_{\mathbf{r}\to\mathbf{r}_c} \d_i\left(G^{(0)}(\mathbf{r},\mathbf{r}_c)-G^{(\mu)}(\mathbf{r},\mathbf{r}_c)\right)u^0,\label{selfforce2}
\end{align}
where $G^{(0)}(\mathbf{r},\mathbf{r}_c)$ and $G^{(\mu)}(\mathbf{r},\mathbf{r}_c)$ are the three-dimensional Green functions for Eqs.~(\ref{Meq2}) and (\ref{Peq}), respectively. Since, due to the reciprocity principle, the Green functions are symmetric with respect to the permutation of their arguments,
\begin{equation}
  G^{(0)}(\mathbf{r},\mathbf{r}_c)=G^{(0)}(\mathbf{r}_c,\mathbf{r}), \quad G^{(\mu)}(\mathbf{r},\mathbf{r}_c)=G^{(\mu)}(\mathbf{r}_c,\mathbf{r}),
\end{equation}
the formula (\ref{selfforce2}) can be transformed into
\begin{gather}
f_i^{\rm sf}=\lim\limits_{\mathbf{r}\to\mathbf{r}_c} \d_i U, \label{selfforce3} \\
U=2\pi q^2\,\left(G^{(0)}(\mathbf{r},\mathbf{r})-G^{(\mu)}(\mathbf{r},\mathbf{r})\right)u^0,\label{selfenergy}
\end{gather}
where the quantity $U$ plays the role of the self-interaction potential energy.  Mention that calculation of the self-interaction energy for the static point charge in the Minkowski spacetime ($u^0=1$) yields
\begin{equation}
U=\lim\limits_{\mathbf{r}\to\mathbf{r}_c} \frac{q A_0}{2}=\frac{\mu q^2}{2},
\end{equation}
and therefore $f^i_{\rm sf}=0$. It means that the second approach implies the averaging procedure automatically.

\section{Self-interaction on a curved spacetime
background: Axial symmetry}\label{axial}

\subsection{General approach}

For the sake of simplicity, let us consider a static axially symmetric spacetime, the metric of which takes the form
\begin{equation}
    ds^2=dt^2-d\rho^2-R(\rho)^2 d\varphi^2 -dz^2.\label{metricax}
\end{equation}
Here $\rho\in[0;+\infty)$ is a radial coordinate, $\varphi\in[0;2\pi]$ is an azimuthal angle, and $R(\rho)$ is a unique unknown function, which will be defined later.

We will assume that a static charge is located at the point
$(\rho_c,\varphi_c,z_c)$ of this spacetime. The Green functions $G^{(0)}(\mathbf{r},\mathbf{r}_c)$ and $G^{(\mu)}(\mathbf{r},\mathbf{r}_c)$  obey the equations
\begin{equation}
    \triangle G^{(0)}(\mathbf{r},\mathbf{r}_c)=-\frac{1}{R(\rho)}\,\delta(\rho-\rho_c)\delta(\varphi-\varphi_c)\delta(z-z_c),\label{Geq1}
\end{equation}
\begin{align}
    \triangle G^{(\mu)}(\mathbf{r},\mathbf{r}_c)&-\mu^2 G^{(\mu)}(\mathbf{r},\mathbf{r}_c)\nonumber\\
    {}&=-\frac{1}{R(\rho)}\,\delta(\rho-\rho_c)\delta(\varphi-\varphi_c)\delta(z-z_c),\label{Geq2}
\end{align}
where $\triangle$ is the Laplace operator relating to the spacetime with the metric (\ref{metricax}):
\begin{equation}
    \triangle \Phi=\frac{1}{R}\d_\rho(R\d_\rho \Phi)+\frac{1}{R^2}\d^2_{\varphi\varphi}\Phi +\d^2_{zz}\Phi.
\end{equation}
The solutions to Eqs.~(\ref{Geq1}), (\ref{Geq2}) can be represented as follows (see, e.g., \cite{SFbook1}, p.~125)
\begin{align}
    & G^{(0)}({\bf r},{\bf r}_c)={}\nonumber\\
    {}&=\frac{1}{2\pi^2}\int\limits_0^\infty d\tau\sum_{n=-\infty}^{\infty} {\rm e}^{in(\varphi-\varphi_c)}\cos\left(\tau\,(z-z_c)\right)g^{(0)}_{n\tau}(\rho,\rho_c),
\end{align}
\begin{align}
    & G^{(\mu)}({\bf r},{\bf r}_c)={}\nonumber\\
    {}&=\frac{1}{2\pi^2}\int\limits_0^\infty d\tau\sum_{n=-\infty}^{\infty} {\rm e}^{in(\varphi-\varphi_c)}\cos\left(\tau\,(z-z_c)\right)g^{(\mu)}_{n\tau}(\rho,\rho_c),
\end{align}
where the radial Green functions $g^{(0)}_{n\tau}$ and $g^{(\mu)}_{n\tau}$ satisfy the equations
\begin{gather}
    {g^{(0)}_{n\tau}}''+\frac{R'}{R}\,{g^{(0)}_{n\tau}}'-\left(\tau^2+\frac{n^2}{R^2}\right)g^{(0)}_{n\tau}=
    -\frac{\delta(\rho-\rho_c)}{R},
\end{gather}
\begin{align}
    {g^{(\mu)}_{n\tau}}''+\frac{R'}{R}\,{g^{(\mu)}_{n\tau}}'&{}-\left(\tau^2+\mu^2+\frac{n^2}{R^2}\right)g^{(\mu)}_{n\tau}\nonumber\\
    {}&{}=-\frac{\delta(\rho-\rho_c)}{R},
\end{align}
respectively, and the prime denotes the derivative with respect to $\rho$. These functions can be written in the following form (for $g_{n\tau}^{(0)}$ we should put $\mu=0$),
\begin{equation}
    g_{n\tau}^{(\mu)}(\rho,\rho_c)=\left\{\begin{array}{l}
      y_{n\tau}^{1(\mu)}(\rho)y_{n\tau}^{2(\mu)}(\rho_c),\ \rho<\rho_c, \\
      y_{n\tau}^{1(\mu)}(\rho_c)y_{n\tau}^{2(\mu)}(\rho),\ \rho>\rho_c,\\
    \end{array}\right.
\end{equation}
where $y_{n\tau}^{1(\mu)}(\rho)$ and $y_{n\tau}^{2(\mu)}(\rho)$ are the solutions of
the corresponding homogeneous equation
\begin{equation}
    y_{n\tau}''+\frac{R'}{R}\,y_{n\tau}'-\left(\tau^2+\mu^2+\frac{n^2}{R^2}\right)\,y_{n\tau}=0,\label{eqhom}
\end{equation}
which are regular at $\rho=0$ and at $\rho=\infty$, respectively, and normalized by their Wronskian,
\begin{equation}
    {\cal W}\left(y_{n\tau}^{1(\mu)}(\rho),y_{n\tau}^{2(\mu)}(\rho)\right)=-\frac{1}{R(\rho)}.\label{eqW}
\end{equation}

Now, to derive the formula for the self-interaction energy we can apply Eq.~(\ref{selfenergy}). Since $u^0=1$ for the metric (\ref{metricax}), in the limit case ${\bf r}\to{\bf r}_c$ (i.e., $\rho\to \rho_c$,
$\varphi\to\varphi_c$, $z\to z_c$), we obtain
\begin{align}
    U(\rho_c)&=\frac{q^2}{\pi}\int\limits_0^\infty d\tau\sum_{n=-\infty}^{\infty}
    \left(y_{n\tau}^{1(0)}(\rho_c)y_{n\tau}^{2(0)}(\rho_c)\right.\nonumber\\
    {}&-\left. y_{n\tau}^{1(\mu)}(\rho_c)y_{n\tau}^{2(\mu)}(\rho_c)\right).\label{selfSU1}
\end{align}
Obviously, this function depends only on the radial coordinate of the charge position $\rho_c$.

\subsection{Cosmic string spacetime}

For further progress we should fix the metric function $R(\rho)$. As an example, we consider the spacetime of a straight infinitely thin cosmic string, for which
\begin{equation}
R(\rho)=b\rho,
\end{equation}
where $b$ is a positive parameter related to a linear mass density of a string. When $b<1$ the density is positive, when $b>1$ it is negative, and in the case $b=1$ we deal with the standard Minkowski spacetime. From a spacetime geometry point of view, this parameter specifies the quantity of the spacetime angular defect.

A set of solutions to the homogeneous equation (\ref{eqhom}), regular at $\rho=0$ ($y_{n\tau}^{1(0)}$ and $y_{n\tau}^{1(\mu)}$) or at $\rho=\infty$ ($y_{n\tau}^{2(0)}$ and $y_{n\tau}^{2(\mu)}$), and satisfying the condition (\ref{eqW}) takes the form
\begin{align}
y_{n\tau}^{1(0)}(\rho)&=\frac{{\rm I}_{|n|/b}(\tau\rho)}{\sqrt{b}},\\
y_{n\tau}^{1(\mu)}(\rho)&=\frac{{\rm I}_{|n|/b}(\sqrt{\tau^2+\mu^2}\cdot\rho)}{\sqrt{b}},\\
y_{n\tau}^{2(0)}(\rho)&=\frac{{\rm K}_{|n|/b}(\tau\rho)}{\sqrt{b}},\\
y_{n\tau}^{2(\mu)}(\rho)&=\frac{{\rm K}_{|n|/b}(\sqrt{\tau^2+\mu^2}\cdot\rho)}{\sqrt{b}},
\end{align}
where ${\rm I}_\nu(x)$ and ${\rm K}_\nu(x)$ are the modified Bessel functions of the first and second kind, respectively. Substituting these expressions into (\ref{selfSU1}), we obtain
\begin{align}
U(\rho)&=\frac{q^2}{\pi b}\int\limits_0^\infty d\tau\sum\limits_{n=-\infty}^{\infty}\left\{{\rm I}_{|n|/b}(\tau \rho)\,{\rm K}_{|n|/b}(\tau \rho)\right.\nonumber \\
{}&-\left.{\rm I}_{|n|/b}(\sqrt{\tau^2+\mu^2}\cdot\rho)\,{\rm K}_{|n|/b}(\sqrt{\tau^2+\mu^2}\cdot\rho)\right\}.
\end{align}
In order to reduce this formula, as the first step, we apply the integral representation for the product of the Bessel functions (see \cite{Gradshteyn}),
\begin{equation}
  {\rm I}_{\nu}(z){\rm K}_{\nu}(z)=\int\limits_0^\infty dx\,{\rm e}^{-2\nu x}{\rm J}_{0}(2z\sinh x),\label{Id2}
\end{equation}
where ${\rm J}_\nu(x)$ is the Bessel function, with the identity
$$\sum\limits_{n=-\infty}^{\infty}{\rm e}^{-2|n|x/b}=\coth\frac{x}{b}$$
and evaluate the series:
\begin{align}
{}&\sum\limits_{n=-\infty}^{\infty}\left\{{\rm I}_{|n|/b}(\tau \rho)\,{\rm K}_{|n|/b}(\tau \rho)\right.\nonumber\\
{}&\qquad {}-\left.{\rm I}_{|n|/b}(\sqrt{\tau^2+\mu^2}\cdot\rho)\,{\rm K}_{|n|/b}(\sqrt{\tau^2+\mu^2}\cdot\rho)\right\}\nonumber\\
{}&=\int\limits_0^\infty dx\,\left\{{\rm J}_{0}(2\tau\rho\sinh x)\right.\nonumber\\
{}&\qquad {}-\left.{\rm J}_{0}(2\sqrt{\tau^2+\mu^2}\cdot\rho\sinh x)\right\}\coth\frac{x}{b}.\label{37}
\end{align}
The swapping integration and summation in (\ref{37}) is possible, because if $\tau\neq0$ the integral $\int_0^\infty dx f_n(x)$ and the series $\sum_n f_n(x)$, where
\begin{align}
f_n(x)&={\rm e}^{-2|n|x/b}\left\{{\rm J}_{0}(2\tau\rho\sinh x)\right.\nonumber\\
{}&\left.-{\rm J}_{0}(2\sqrt{\tau^2+\mu^2}\cdot\rho\sinh x)\right\},\nonumber
\end{align}
converge uniformly. Since the Bessel function ${\rm J}_0(x)$ at the origin behaves as ${\rm J}_0(x)\approx 1-x^2/4$, the integrand in the right-hand side of (\ref{37}) is finite when $x\to 0$. Now we can integrate this expression over $\tau$. As a result, we obtain
\begin{equation}
U(\rho)=\frac{U_0}{\pi\mu\rho b}\int\limits_0^\infty dx\coth\frac{x}{b}\cdot\frac{1-\cos(2\mu\rho\sinh x)}{\sinh x},\label{Uax1}
\end{equation}
where $U_0=\mu q^2/2$ is the charge field energy for the Minkowski spacetime (\ref{E0}), i.e., when $b=1$.

It is worth noting that in Eq.~(\ref{Uax1}) we have interchanged the order of integration
%(or, for (\ref{37}), integration and summation)
$$\int_0^\infty d\tau\int_0^\infty dx F(\tau,x)= \int_0^\infty dx\int_0^\infty d\tau F(\tau,x),$$ where
\begin{align}
F(\tau,x)&=\left\{{\rm J}_{0}(2\tau\rho\sinh x)-{\rm J}_{0}(2\sqrt{\tau^2+\mu^2}\cdot\rho\sinh x)\right\}\nonumber \\
{}&\times\coth\frac{x}{b}.\nonumber
\end{align}
To justify this transformation, let us consider, firstly, the case when both lower limits are nonzero. The swapping $\int_A^\infty d\tau\int_B^\infty dx = \int_B^\infty dx\int_A^\infty d\tau$ is allowed, because the integrals $\int_A^\infty d\tau F(\tau,x)$ and $\int_B^\infty dx F(\tau,x)$ converge uniformly for $x>B>0$ and $\tau>A>0$, respectively. Second, since at $x\to0$ we have $F(\tau,x)\approx \mu^2\rho^2 b x>0$, there exists a region in the vicinity of the origin (say, $x\in[0,B]$), where the function $F(\tau,x)$ is nonnegative. Due to a theorem from \cite{Zorich} the swapping $\int_0^\infty d\tau\int_0^B dx = \int_0^B dx\int_0^\infty d\tau$ is allowed again. At last, since the integral $\int_0^\infty dx F(\tau,x)$ has an integrable singularity at $\tau=0$, to prove the interchanging $\int_0^A d\tau\int_0^\infty dx = \int_0^\infty dx\int_0^A d\tau$ we should change a variable $\tau\to 1/\tau$. Obtained integrals converge uniformly as well.

From Eq.~(\ref{Uax1}), one can see that the self-interaction potential energy depends on the dimensionless parameter $\mu \rho$ only. In order to demonstrate this dependence in the explicit form, we should consider the integral in (\ref{Uax1}) on the complex plane,
\begin{equation}
U(\rho)=\frac{U_0}{\pi\mu\rho b}\Re\int\limits_0^\infty dz\coth\frac{z}{b}\cdot\frac{1-{\rm e}^{2i\mu\rho\sinh z}}{\sinh z},\label{Uax2}
\end{equation}
and change the path of integration. The new contour consists of two parts: first, the half-line $\Im z=\pi/2,\ \Re z\leq 0$, and second, the segment of the imaginary axis $\Im z\in(0;\pi/2)$; meanwhile, the poles $z=\pi b n i$ of the function $\coth z/b$ ($n$ is integer and less than $1/2b$) and the point $z=0$ should be excluded.
%, i.e., for the Minkowski spacetime, the formula (\ref{Uax1}) gives the usual quantity $U_0=\mu q^2/2$ (see %\ref{E0})
%$$U(\rho)=\frac{q^2}{2\pi\rho}\int\limits_0^\infty dx\coth x\cdot\frac{1-\cos(2\mu\rho\sinh x)}{\sinh x}=\frac{\mu %q^2}{2}.$$

\begin{figure}[b]
\includegraphics[height=6cm]{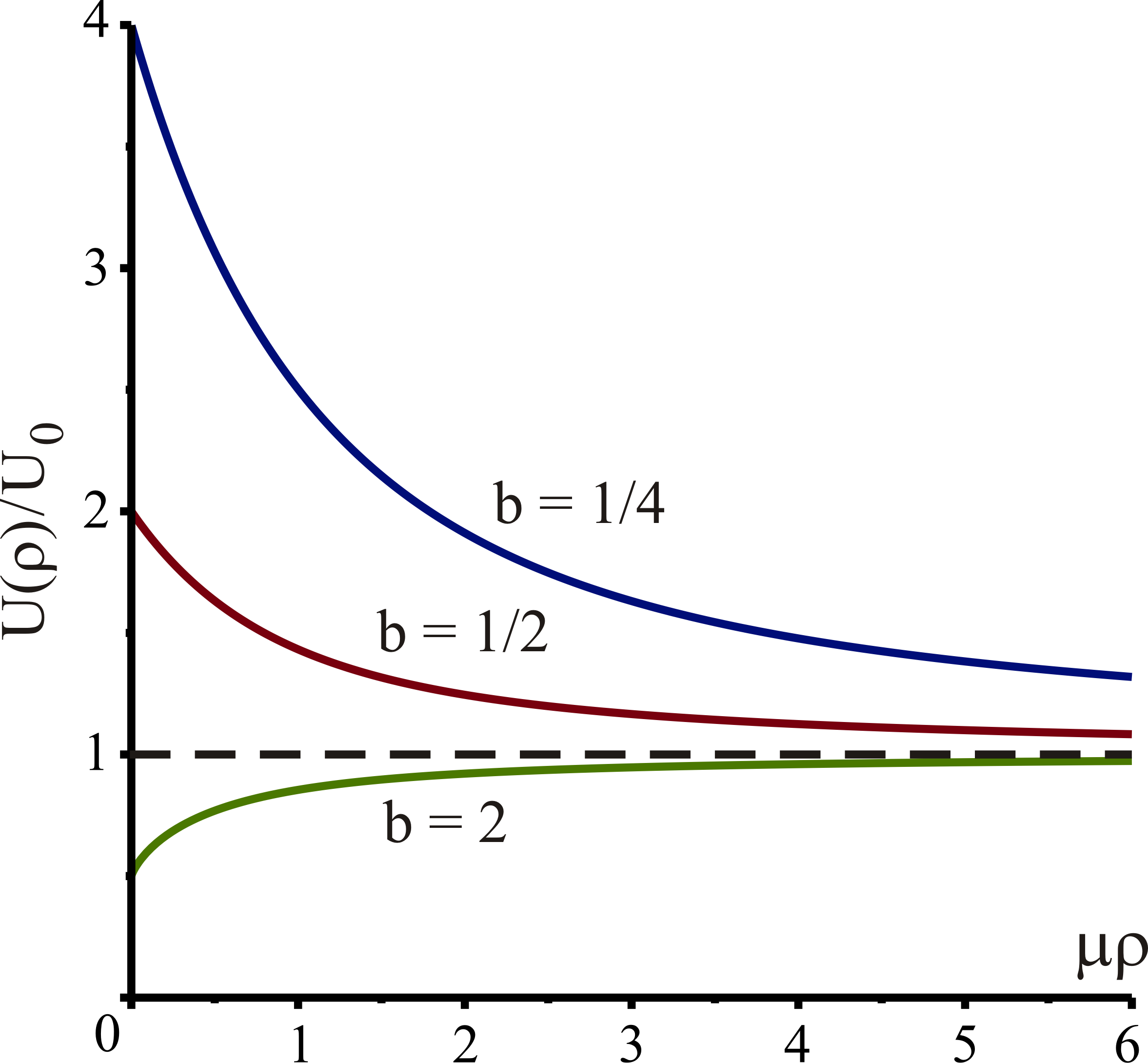} \caption{Plots of the self-interaction potential energy $U(\rho)$ normalized on $U_0=\mu q^2/2$. When $b>1$, $U(\rho)$ is an increasing function, and the self-force attracts the charge to the cosmic string, while, when $b<1$,
$U(\rho)$ is a decreasing function, therefore the charge and the string repel one another.}\label{String}
\end{figure}

After some routine calculations we obtain (\ref{Uax1}) in the new, more convenient form,
\begin{align}
\frac{U(\rho)}{U_0}&=1+\frac{1}{\mu \rho}\sum\limits_{n=1}^{[1/2b]}\frac{1-{\rm e}^{-2\mu\rho\sin \pi b n}}{\sin \pi b n}\nonumber\\
{}&-\frac{\sin \frac{\pi}{b}}{\pi b\mu\rho}\int\limits_0^\infty \frac{\left(1-{\rm e}^{-2\mu\rho\cosh x}\right)}{(\cosh \frac{2x}{b}-\cos \frac{\pi}{b})}\frac{dx}{\cosh x}.\label{Uax3}
\end{align}
This relation generalizes the corresponding formulas from \cite{Khus95} and \cite{LinetA} for the BP electrodynamics. If the angular defect parameter $b=1/(2N+1)$, where $N$ is nonnegative integer, the last term in (\ref{Uax3}) vanishes and this formula reduces to
\begin{align}
\frac{U(\rho)}{U_0}&=1+\frac{1}{\mu \rho}\sum\limits_{n=1}^{N}\frac{1-{\rm e}^{-2\mu\rho\sin \frac{\pi n}{2N+1}}}{\sin \frac{\pi n}{2N+1}}.\label{UaxRed1}
\end{align}
In contrast, if $b=1/(2N)$ the last term tends to $\frac{1}{2\mu\rho}(1-\exp(-2\mu\rho))$, and the formula (\ref{Uax3}) transforms to
\begin{align}
\frac{U(\rho)}{U_0}&=1+\frac{1}{\mu \rho}\sum\limits_{n=1}^{N-1}\frac{1-{\rm e}^{-2\mu\rho\sin \frac{\pi n}{2N}}}{\sin \frac{\pi n}{2N}}+\frac{1-{\rm e}^{-2\mu\rho}}{2\mu \rho}.\label{UaxRed2}
\end{align}
When $\rho\to\infty$ the function $U(\rho)$ obviously tends to $U_0$, while at the origin $U(0)=U_0/b$. Plots of the function $U(\rho)$ for some values of the parameter $b$ are presented in Fig.~\ref{String}.
Let us consider the formula (\ref{Uax3}) in two limiting cases. First, when the angular defect (or the string mass density) is very small, i.e., for $|1-b|\ll1$, Eq.~(\ref{Uax3}) yields
\begin{gather}
    \frac{U(\rho)}{U_0}\approx 1+\frac{(1-b)}{2\mu \rho}\left\{\frac{\pi}{4}-\int\limits_0^\infty\frac{{\rm e}^{-2\mu\rho\cosh x}dx}{\cosh^3x}
    \right\}.
\end{gather}
Second, when $\mu \rho\gg 1$ the asymptotic behavior of $U(\rho)/U_0$ is determined by the expression
\begin{equation}
    \frac{U(\rho)}{U_0}\approx 1+\frac{1}{\mu\rho}P(b)-L(\mu\rho,b),
\end{equation}
where
\begin{align}
P(b)&=\sum\limits_{n=1}^{[1/2b]}\frac{1}{\sin \pi b n}-\frac{\sin \frac{\pi}{b}}{\pi b}\int\limits_0^\infty \frac{1}{(\cosh \frac{2x}{b}-\cos \frac{\pi}{b})}\frac{dx}{\cosh x}\nonumber\\
{}&=\frac{1}{\pi}\int\limits_0^\infty \frac{dx}{\sinh x}\left(\frac{1}{b}\coth\frac{x}{b}-\coth x\right),\label{UaxP}
\end{align}
and the remainder term $L(\mu\rho,b)$ behaves as follows:
\begin{equation}
L(\mu\rho,b)\sim\left\{\begin{array}{l} \dfrac{|\cot\frac{\pi}{2b}|}{2b\sqrt{\pi (\mu\rho)^3}}\,{\rm e}^{-2\mu\rho},\quad \hbox{when}\ b>\frac12,\medbreak\\ \frac{1}{2\mu\rho}\,{\rm e}^{-2\mu\rho},\quad \hbox{when}\ b=\frac12,\medbreak\\  \dfrac{1}{\mu\rho\sin\pi b}\,{\rm e}^{-2\mu\rho\sin\pi b},\quad \hbox{when}\ b<\frac12.\end{array}\right.\label{remainder-ax}
\end{equation}
The term $P(b)/(\mu \rho)$ in the second form is a standard {\it renormalized} expression for the
self-interaction potential energy \cite{SFstring} in the Maxwell model. In the framework of the BP electrodynamics, we derive it as the limiting case of the general formula (\ref{Uax3}) without any regularization procedure.

Analyzing the obtained function $U(\rho)$ for the self-interaction energy, we can conclude that if $b>1$, the function $U(\rho)$ is an increasing one, and the self-force attracts the charge to the cosmic string. In contrast, when $b<1$ the function $U(\rho)$ is a decreasing one, and in this case repulsion between the charge and the string occurs.

\section{Self-interaction on a curved spacetime
background: Spherical symmetry}\label{spherical}

\subsection{General approach}

Let us proceed to the case, when the spacetime is static and it has the spherical symmetry. For the sake of simplicity, the metric of this spacetime takes the form
\begin{equation}
    ds^2=dt^2-dr^2-R(r)^2\left(d\theta^2+\sin^2\theta
    d\varphi^2\right),\label{metricsph}
\end{equation}
where $r\in[0;+\infty)$ is a radial coordinate, $\theta\in [0;\pi]$ and
$\varphi\in[0;2\pi]$ are angular coordinates of the spherical system. $R(r)$ is a unique undefined metric function.

We will assume that a static charge is placed at the
point ($r_c$, $\theta_c$, $\varphi_c$) of the spacetime. In order to derive the expression for the self-interaction potential energy we use the same scheme as it was demonstrated above. The Green functions
$G^{(0)}(\mathbf{r},\mathbf{r}_c)$ and $G^{(\mu)}(\mathbf{r},\mathbf{r}_c)$ satisfy the equations
\begin{equation}
    \triangle G^{(0)}(\mathbf{r},\mathbf{r}_c)=-\frac{\delta(r-r_c)\delta(\theta-\theta_c)
    \delta(\varphi-\varphi_c)}{R(r)^2\sin\theta},\label{Geq3}
\end{equation}
\begin{align}
    \triangle G^{(\mu)}(\mathbf{r},\mathbf{r}_c)&-\mu^2 G^{(\mu)}(\mathbf{r},\mathbf{r}_c)\nonumber\\
    {}&=-\frac{\delta(r-r_c)\delta(\theta-\theta_c)
    \delta(\varphi-\varphi_c)}{R(r)^2\sin\theta}.\label{Geq4}
\end{align}
The Laplacian relating to the spacetime metric (\ref{metricsph}) takes the form
\begin{align}
    \triangle \Phi&=\frac{1}{R^2}\d_r(R^2\d_r \Phi)+\frac{1}{R^2\sin\theta}\d_{\theta}(\sin\theta \d_\theta\Phi)\nonumber\\
    {}&+\frac{1}{R^2\sin^2\theta}\d^2_{\varphi\varphi}\Phi.
\end{align}
The solutions to Eqs.~(\ref{Geq3}), (\ref{Geq4}) can be represented as follows (see, e.g., \cite{SFbook1}, p.~120):
\begin{equation}
    G^{(\mu)}({\bf r},{\bf r}_c)=\sum_{n=0}^{\infty}\sum_{m=-n}^{n}
    g_n^{(0)}(r,r_c) {\rm Y}_{nm}(\theta,\varphi){\rm Y}^*_{nm}(\theta_c,\varphi_c),
\end{equation}
\begin{equation}
    G^{(\mu)}({\bf r},{\bf r}_c)=\sum_{n=0}^{\infty}\sum_{m=-n}^{n}
    g_n^{(\mu)}(r,r_c) {\rm Y}_{nm}(\theta,\varphi){\rm Y}^*_{nm}(\theta_c,\varphi_c),
\end{equation}
where the radial Green functions $g^{(0)}_{n}$ and $g^{(\mu)}_{n}$ satisfy the equations
\begin{gather}
    {g^{(0)}_{n}}''+\frac{2R'}{R}\,{g^{(0)}_{n}}'-\frac{n(n+1)}{R^2} g^{(0)}_{n}=
   -\frac{\delta(r-r_c)}{R^2},
\end{gather}
\begin{align}
    {g^{(\mu)}_{n}}''&+\frac{2R'}{R}\,{g^{(\mu)}_{n}}'-\left(\mu^2+\frac{n(n+1)}{R^2}\right)g^{(\mu)}_{n}\nonumber\\
    {}&=-\frac{\delta(r-r_c)}{R^2},
\end{align}
respectively, ${\rm Y}_{nm}(\theta,\varphi)$ is the spherical function, and the asterisk denotes complex conjugation. In contrast with the previous section, here the prime relates to the derivative with respect to $r$.

These functions can be written in the following form (for $g_{n}^{(0)}$ we should put $\mu=0$),
\begin{equation}
    g_{n}^{(\mu)}(r,r_c)=\left\{\begin{array}{l}
      y_{n}^{1(\mu)}(r)y_{n}^{2(\mu)}(r_c),\ r<r_c, \\
      y_{n}^{1(\mu)}(r_c)y_{n}^{2(\mu)}(r),\ r>r_c,\\
    \end{array}\right.
\end{equation}
where $y_{n}^{1(\mu)}(r)$ and $y_{n}^{2(\mu)}(r)$ are the solutions of
the corresponding homogeneous equation
\begin{equation}
    y_{n}''+\frac{2R'}{R}\,y_{n}'-\left(\mu^2+\frac{n(n+1)}{R^2}\right)\,y_{n}=0,\label{eqhom2}
\end{equation}
which are regular at $r=0$ and at $r=\infty$, respectively, and normalized by their Wronskian
\begin{equation}
    {\cal W}\left(y_{n}^{1(\mu)}(r),y_{n}^{2(\mu)}(r)\right)=-\frac{1}{R(r)^2}.\label{eqW2}
\end{equation}

For the metric (\ref{metricsph}) we have $u^0=1$ (see (\ref{selfenergy})) and the self-interaction energy formula gives \begin{align}
    U(r_c)&=\frac{q^2}{2}\sum_{n=0}^\infty
    (2n+1)\left(y_n^{1(0)}(r_c)y_n^{2(0)}(r_c)\right.\nonumber\\
    {}&-\left.y_n^{1(\mu)}(r_c)y_n^{2(\mu)}(r_c)\right).\label{selfMU1}
\end{align}
It is easy to see this function depends only on the radial coordinate of the charge position $r_c$.

\subsection{Global monopole spacetime}

Let us fix again the metric function $R(r)$. As an example we consider the spacetime
of a global monopole spacetime, for which
\begin{equation}
R(r) = br,
\end{equation}
where $b$ is a positive parameter related to the scale characterizing spontaneous
symmetry breaking. In the case $b = 1$ we deal again with the standard Minkowski spacetime. When $b\neq 1$ this spacetime possesses the conic singularity at the origin and
it is not locally flat. From this point of view, the parameter $b$ determines a defect of solid angle.

In this case, solutions to the homogeneous equation (\ref{eqhom2}), regular at $r=0$ ($y_{n}^{1(0)}$ and $y_{n}^{1(\mu)}$) or at $r=\infty$ ($y_{n}^{2(0)}$ and $y_{n}^{2(\mu)}$), and obeying the condition (\ref{eqW2}) take the form
\begin{align}
y_{n}^{1(0)}(r)&=\frac{r^{N_n(b)/2b}}{\sqrt{N_n(b)br}},\\
y_{n}^{1(\mu)}(r)&=\frac{{\rm I}_{\frac{N_n(b)}{2b}}(\mu r)}{b\sqrt{r}},\\
y_{n}^{2(0)}(r)&=\frac{r^{-N_n(b)/2b}}{\sqrt{N_n(b)br}},\\
y_{n}^{2(\mu)}(r)&=\frac{{\rm K}_{\frac{N_n(b)}{2b}}(\mu r)}{b\sqrt{r}},
\end{align}
where $N_n(b)=\sqrt{(2n+1)^2+b^2-1}$.
Substituting these functions into (\ref{selfMU1}), we obtain the self-interaction energy expression (normalized to $U_0$)
\begin{align}
\frac{U(r)}{U_0}&=\frac{1}{\mu rb}\sum_{n=0}^\infty\frac{(2n+1)}{N_n(b)}\nonumber\\
{}&\times\left\{1-\frac{N_n(b)}{b}\,{\rm
I}_{\frac{N_n(b)}{2b}}(\mu r)\,{\rm K}_{\frac{N_n(b)}{2b}}(\mu r)\right\}.\label{Usph1}
\end{align}
Applying the identity \cite{Gradshteyn}
\begin{equation}
1-2\nu {\rm I}_{\nu}(x){\rm K}_{\nu}(x)=
     2x\int\limits_0^\infty dt\,{\rm e}^{-2\nu t}\cosh t {\rm J}_{1}(2x\sinh t),
\end{equation}
we can transform (\ref{Usph1}) into the second form
\begin{align}
\frac{U(r)}{U_0}&=\frac{2}{b}\int\limits_0^\infty dt\, \cosh t\cdot {\rm
J}_1(2\mu r\sinh
t)\nonumber \\
{}&\times\left\{\sum_{n=0}^\infty\frac{(2n+1)}{N_n(b)}\exp\left(-\frac{N_n(b)t}{b}\right)\right\}.\label{Usph2}
\end{align}

It is worth noting again that in Eq.~(\ref{Usph2}) we have switched integration and summation,
%(or, for (\ref{37}), integration and summation)
$$\int\limits_0^\infty dt\sum\limits_{n=0}^\infty F_n(t)= \sum\limits_{n=0}^\infty\int\limits_0^\infty dt F_n(t),$$ where
\begin{align}
F_n(t)&=\cosh t\, {\rm J}_1(2\mu r\sinh
t)\frac{(2n+1)}{N_n(b)}\exp\left(-\frac{N_n(b)t}{b}\right).\nonumber
\end{align}
First, the swapping $\int_A^\infty dt \sum_{n} = \sum_n\int_A^\infty dt$ is allowed, because the integral $\int_A^\infty dt F_n(t)$ and the series $\sum_n F_n(t)$ converge uniformly for $n\geq 0$ and $t>A>0$, respectively. Second, since at $x\to0$ we have $F_n(t)>0$, there exists a region in the vicinity of the origin (say, $t\in[0,A]$), where the function $F_n(t)$ is nonnegative and the swapping $\int_0^A dt \sum_{n} = \sum_{n} \int_0^A dt$ is allowed as well.

When $b=1$ one has to obtain $U(r)=U_0$. Indeed, in this case we have $N_n(1)=2n+1$,
$$\sum_{n=0}^\infty\frac{(2n+1)}{N_n(1)}\exp\left(-{N_n(1)t}\right)=\sum_{n=0}^\infty {\rm e}^{-(2n+1)t}=\frac{1}{2\sinh t},$$
and therefore
\begin{equation}
\frac{U(r)}{U_0}=\int\limits_0^\infty dt\, \frac{\cosh t}{\sinh t}\cdot {\rm
J}_1(2\mu r\sinh
t)=1.
\end{equation}
As a result, we prove the useful identity, which is valid for any positive $x$,
\begin{equation}
\sum_{n=0}^\infty \left\{1-(2n+1) {\rm I}_{n+1/2}(x){\rm K}_{n+1/2}(x)\right\}=x.\label{Id1}
\end{equation}
Now, substituting (\ref{Id1}) for $x=\mu r b$ into (\ref{Usph1}), we can reconstruct the formula for $U(r)$
in a form that is more convenient to analyze:
\begin{align}
\frac{U(r)}{U_0}&=1+\frac{1}{\mu rb}\sum_{n=0}^{\infty}\left(\frac{2n+1}{N_n(b)}-1\right)\nonumber\\
&{}+\frac{1}{\mu rb}\sum_{n=0}^\infty(2n+1)\left\{{\rm
I}_{n+\frac12}(\mu rb)\,{\rm K}_{n+\frac12}(\mu r b)\right.\nonumber\\
{}&-\left.\frac{1}{b}\,{\rm
I}_{\frac{N_n(b)}{2b}}(\mu r)\,{\rm K}_{\frac{N_n(b)}{2b}}(\mu r)\right\}.\label{Usph3}
\end{align}

\begin{figure}[b]
\includegraphics[height=6cm]{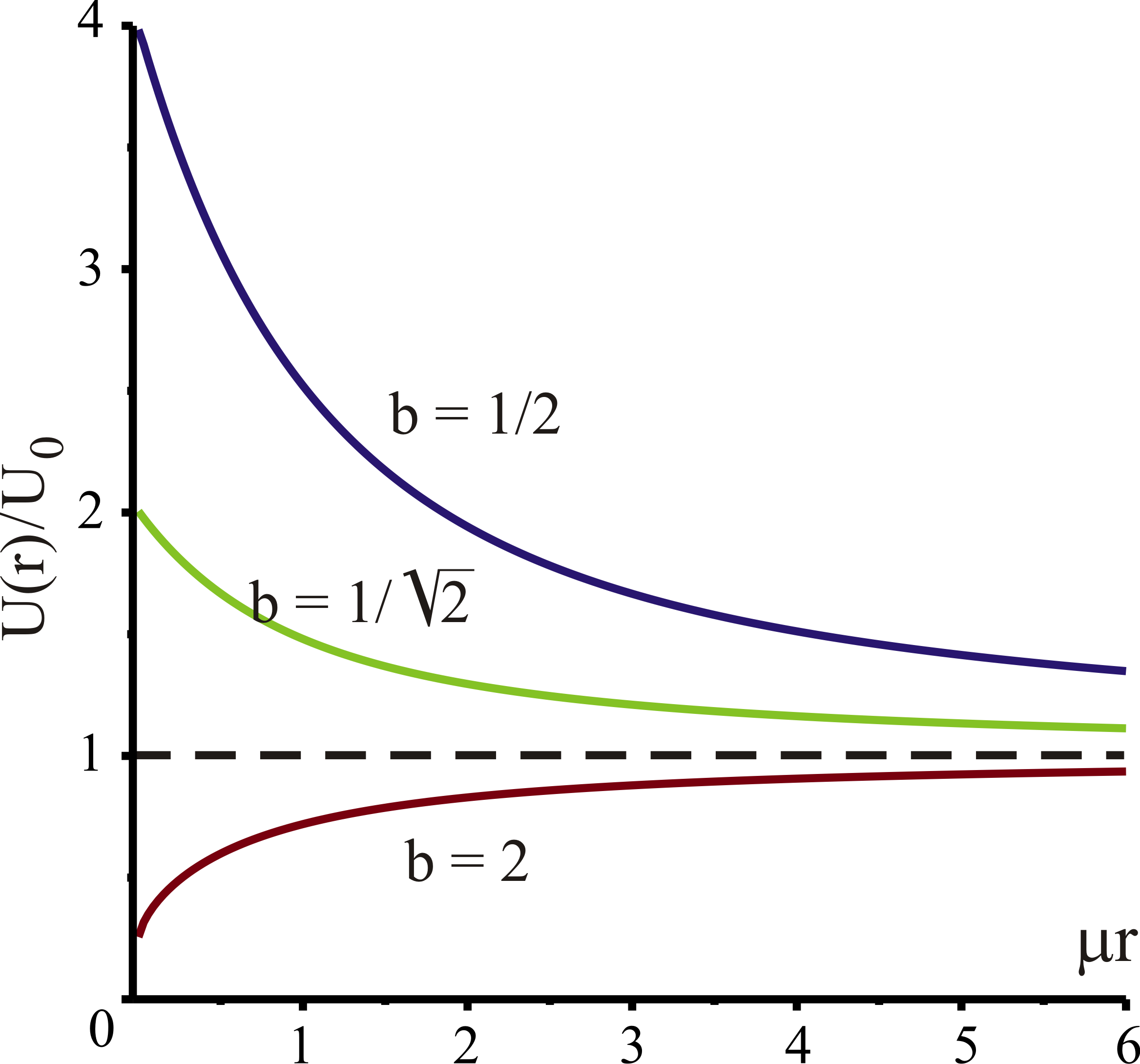}\caption{Plots of the self-interaction potential energy $U(r)$ normalized on $U_0=\mu q^2/2$ in the case of the static charge on the global monopole spacetime background. When the angular defect parameter $b>1$, $U(r)$ is an increasing function, and the self-force attracts the charge to the monopole, while, when $b<1$, $U(r)$ is a decreasing function, therefore the charge and the monopole repel one another.}\label{Mon}
\end{figure}

Due to the representation (\ref{Usph1}) one can derive the value of $U(0)$ at the origin. Since the function $1-2\nu {\rm I}_\nu (x){\rm K}_\nu(x)$ when $x\to 0$ behaves as
\begin{equation}
1-2\nu {\rm I}_\nu(x){\rm K}_\nu(x)\sim \left[\begin{array}{l}                                                             \displaystyle\left(\frac{x}{2}\right)^{2\nu}\frac{\Gamma(1-\nu)}{\Gamma(1+\nu)},\ \hbox{if }\nu<1,\medbreak\\
\displaystyle\frac{x^2}{2}\ln\frac{2}{x},\ \hbox{if }\nu=1,\medbreak\\
\displaystyle\frac{x^2}{2(\nu^2-1)},\ \hbox{if }\nu>1, \end{array}\right.
\end{equation}
the leading order in (\ref{Usph1}) corresponds to the zeroth term of the series, i.e.,
\begin{equation}
\frac{U(r)}{U_0}\sim \frac{1-{\rm I}_{1/2}(\mu r){\rm K}_{1/2}(\mu r)}{\mu r b^2}\sim \frac{1}{b^2}.
\end{equation}
Thus, the self-interaction energy is finite at the origin as well as in the case of the cosmic string spacetime. But now this value is $U(0)=U_0/b^2$. In the general case, we can apply numerical methods to the plot of the function $U(r)$. The curves corresponding to various values of the parameter $b$ are presented in Fig.~\ref{Mon}. Analyzing them, we can conclude that if $b>1$, the function $U(r)$ is an increasing one, and the self-force attracts the charge to the monopole. In contrast, when $b<1$ the function $U(r)$ is a decreasing one, therefore the charge and the monopole repel one another.

When the angular defect is very small, i.e., for $|1-b| \ll 1$, we have
\begin{equation}
\sum_{n=0}^{\infty}\left(\frac{2n+1}{N_n(b)}-1\right)\sim \frac{\pi^2(1-b)}{8},
\end{equation}
\begin{align}
{}&\sum_{n=0}^\infty(2n+1)\left\{{\rm
I}_{n+\frac12}(\mu rb)\,{\rm K}_{n+\frac12}(\mu r b)\right.\nonumber\\
{}&-\left.\frac{1}{b}\,{\rm
I}_{\frac{N_n(b)}{2b}}(\mu r)\,{\rm K}_{\frac{N_n(b)}{2b}}(\mu r)\right\}\nonumber\\
{}&\sim (b-1)\int\limits_0^\infty dt {\rm J}_0(2\mu r\sinh t)\left(\frac{\cosh t}{\sinh^2 t}-\frac{t}{\sinh^3
    t}\right),
\end{align}
where for the latter expression the formula (\ref{Id2}) has been applied.
As a result, the representation (\ref{Usph3}) gives
\begin{align}
    \frac{U(r)}{U_0}&\approx 1+(1-b)F(\mu r),\nonumber\\
    F(x)&=\frac{1}{x}\left\{\frac{\pi^2}{8}-
    \int\limits_0^\infty dt\, {\rm J}_0(2x\sinh t)\right.\nonumber\\
    {}&\left.\times\left(\frac{\cosh t}{\sinh^2 t}-\frac{t}{\sinh^3
    t}\right)\right\}.%\\
    %&{}\approx 1+\frac{(1-b)}{\mu r}\left\{\frac{\pi^2}{8}-
    %\frac12\sum_{k=0}^\infty \frac{(2k)!}{(2\mu r)^{2k+1}}\left(
%\frac{1}{(2k+1)}+\frac{1}{(2k+3)}\right)\right\}.
\end{align}
The function $F(x)$ is a monotonically decreasing one. At the origin $F(0)=2$, while at infinity
$F(x)$ behaves according to the asymptotic expansion
\begin{equation}
F(x)\approx\frac{\pi^2}{8x}-
    \sum_{k=1}^\infty \frac{(2k-2)!}{(2x)^{2k}}\left(\frac{1}{(2k-1)}+\frac{1}{(2k+1)}\right).
\end{equation}

In order to derive the asymptotic formula when $\mu r\gg 1$ for a generic value of $b$, we should apply the
Abel-Plana summation method to the last term in (\ref{Usph3}). In this way, for $b>1$ we can rewrite the series
as follows (here we have denoted $a^2\equiv b^2-1$):
\begin{widetext}
\begin{align}
{}&\sum_{n=0}^\infty(2n+1)\left\{{\rm
I}_{n+\frac12}(\mu rb)\,{\rm K}_{n+\frac12}(\mu r b)-\frac{1}{b}\,{\rm
I}_{\frac{N_n(b)}{2b}}(\mu r)\,{\rm K}_{\frac{N_n(b)}{2b}}(\mu r)\right\}\nonumber\\
{}&{}= b\int\limits_0^\infty dx\,x\,\Re {\rm I}_{ix/2}(\mu r)\,{\rm K}_{ix/2}(\mu r)\left[\frac{1}{\exp(\pi x)+1}-\frac{1}{\exp(\pi\sqrt{b^2x^2+a^2})+1}\right]\nonumber\\
{}&{}+\frac{a^2}{2b}\int\limits_0^{1}x\,{\rm I}_{ax/2b}(\mu r){\rm K}_{ax/2b}(\mu r)\tanh\left(\frac{\pi a\sqrt{1-x^2}}{2}\right)\,dx .
\end{align}
\end{widetext}
When $b<1$ the analytic continuation of this expression should be done. Further, substituting the asymptotic expansion
for the product of the Bessel functions \cite{AbrSteg},
\begin{align}
{\rm I}_\frac{\nu}{2}(z){\rm K}_\frac{\nu}{2}(z)&\approx \frac{1}{2z}-\frac{1}{2}\frac{\nu^2-1}{(2z)^3}\nonumber\\
{}&+\frac{1\cdot 3}{2\cdot 4}\frac{(\nu^2-1)(\nu^2-9)}{(2z)^5}+\dots ,
\end{align}
we obtain
\begin{align}
    \frac{U(r)}{U_0}&=1+\frac{1}{\mu rb}\sum_{n=0}^{\infty}\left(\frac{2n+1}{N_n(b)}-1\right)
    -\frac{(1-b^2)}{6(\mu rb)^2}\nonumber \\
    {}&-\frac{(1-b^4)}{60(\mu rb)^4}-\frac{(4+21b^2-25b^6)}{840(\mu rb)^6}+\dots.\label{UsphExp}
\end{align}
Here the second term is a standard renormalized expression for the
self-interacting energy in the Maxwell model \cite{SFmonopole}. The next term, which can be indicated as essentially a BP one, decreases as $r^{-2}$, in contrast with the previous, axially symmetric case.

\section{Conclusion}\label{conclusion}

In this work, we have considered the higher derivative
Bopp-Podolsky electrodynamics on the curved spacetime background. In the
framework of this model, we have studied the self-interaction potential energy
of the static charge for two specific cases. In the first case, the background is the static axially symmetric
spacetime of a straight thin cosmic string defined by the metric (\ref{metricax}), while the second case relates to the global gravitating monopole. The latter spacetime is a static spherically symmetric one with the metric (\ref{metricsph}).

For both configurations we have obtained exact expressions for the self-interaction energy (see (\ref{Uax1}), (\ref{Uax3}) for the cosmic string case and (\ref{Usph1}), (\ref{Usph2}), (\ref{Usph3}) for the monopole case), which have been written down using integral and series representations and have been depicted in Figs.~\ref{String} and \ref{Mon}. Let us emphasize some properties of these expressions:
\begin{enumerate}
\item Due to the symmetry of the spacetimes under consideration, self-interaction energy $U$ depends on a radial coordinate only.
\item Both of these spacetimes possess angular defects and therefore they have singularities at the origin. In spite of this, the self-interaction energy expressions appear to be finite everywhere, even on the spacetime singularity. The self-force expression remains ill-defined at the origin, because the function $U$ is not smooth there.
\item At infinity the energy value tends to $U_0=\mu q^2/2$, i.e., to the value of the self-interaction energy in the Minkowski spacetime.
\item When the angular defect parameter $b>1$, the self-interaction energy $U$ is an increasing function of a radial coordinate, and the self-force attracts the charge to the string or to the monopole, respectively. When $b<1$, $U$ is a decreasing function, therefore the charge and the string/monopole repulse one another. If $b=1$ we have no topological defects and the spacetime returns to the Minkowski one. In this case, the self-interaction is constant and equal to $U_0$.
\item The parameter $\mu$ of the Bopp-Podolsky model is included into formulas for the normalized self-interaction energy $U/U_0$ only as a scale factor in front of the radial coordinate (see (\ref{Uax2}), (\ref{Uax3}) and (\ref{Usph1}), (\ref{Usph2}), (\ref{Usph3})).
\item The formal limit $U_{\rm reg}=\lim_{\mu\to\infty}(U-U_0)$ reproduces the standard results for the regularized self-interaction energy in the framework of the Maxwell electrodynamics (see \cite{SFstring} and \cite{SFmonopole}):
    $$U_{\rm reg}(\rho)=\frac{q^2}{2\pi\rho}\int\limits_0^\infty \frac{dx}{\sinh x}\left(\frac{1}{b}\coth\frac{x}{b}-\coth x\right)$$
    for the cosmic string spacetime and
    $$U_{\rm reg}(r)=\frac{q^2}{2rb}\sum_{n=0}^{\infty}\left(\frac{2n+1}{\sqrt{(2n+1)^2+b^2-1}}-1\right)$$
    for the global monopole spacetime, respectively.
\item For the case of the global monopole, the remainder term $U-U_0-U_{\rm reg}$, which can be indicated as essentially a Bopp-Podolsky one, behaves as $r^{-2}$ (see (\ref{UsphExp})). It decreases very slowly in comparison with the cosmic string case, for which the remainder term drops exponentially (see (\ref{remainder-ax})).
\end{enumerate}
Thus, these examples examined in
the present paper demonstrate that the self-interaction energy is regular for the static point charge
in curved spacetimes, at least, for spacetimes with a simple casual structure.
It certainly would be interesting to apply our approach to a spacetime with horizons, for instance, to the Schwarzschild spacetime, and generalize the results obtained in \cite{SchSF}.

%For the configuration with the cosmic string

%This spacetime is locally flat but it possesses a
%delta-like singularity at the string.

%static spherically
%symmetric configurations with the Yang-Mills field possessing the SU(2) symmetry. Based on the Wu-Yang ansatz
%for the gauge field, we have obtained a three-parameter
%family of the explicit exact solutions to the nonlinear
%Einstein-Yang-Mills equations. Only one solution from
%this family is regular and belongs to the class of wormhole
%spacetimes. We have denoted this solution as a nonminimal
%Wu-Yang wormhole [see Eq. (40)].

\appendix

\acknowledgments
The author thanks V.~Perlick for the fruitful discussion.
This work was supported by the Program of Competitive Growth of KFU
(Project No.~0615/06.15.02302.034), and by Russian Foundation for Basic Research
(Grant RFBR No.~14-02-00598).

\end{document}